 \documentclass[journal=jpccck]{achemso}
\usepackage{color}
\usepackage{verbatim}
\usepackage{mciteplus}
\usepackage{amsbsy}
\usepackage{amstext}
\usepackage{amssymb}
\usepackage{natmove}
\usepackage{caption}
\usepackage{float}
\usepackage{geometry}
\usepackage{setspace}
\usepackage{xkeyval}
\usepackage{graphicx}
\usepackage{subfig}
\usepackage{placeins}
\usepackage{longtable}
\usepackage{notes2bib}
\usepackage{natbib}
\usepackage{colortbl}

\DeclareGraphicsExtensions{.jpg, .eps, .png}

\definecolor{JPCCBlue}{RGB}{34,80,169}
\definecolor{NLRed}{RGB}{215,24,30}
\definecolor{abstractcolor}{RGB}{255,243,201}
\definecolor{tablegray}{gray}{0.85}

\newcolumntype{M}{>{\centering\arraybackslash}m{2.cm}}

 \usepackage[version=3]{mhchem} % Formula subscripts using \ce{}
 \usepackage[unicode=true,
  bookmarks=true,bookmarksnumbered=false,bookmarksopen=false,
  breaklinks=false,pdfborder={0 0 0},backref=false,colorlinks=true]
  {hyperref}
 \hypersetup{pdftitle={Remarkable Hydrogen Storage on Beryllium Oxide Clusters: First Principles Calculations},
  pdfauthor={Ravindra Shinde, Meenakshi Tayade},
  pdfsubject={Computational Condensed Matter},
  unicode=false,pdftoolbar=true,pdfmenubar=true,pdffitwindow=true,pdfstartview={FitH},pdfnewwindow=true,pdfcreator={RevTeX},linkcolor=red,citecolor=blue,urlcolor=blue}
 
 \makeatletter

\usepackage{titlesec}
\titleformat{\section}{\bfseries\sffamily\color{JPCCBlue}}{\thesection.~}{0pt}{}
\titleformat{\subsection}[runin]{\bfseries\sffamily\normalsize}{\indent\thesubsection.~}{0pt}{}[.]
\titlespacing{\subsection}{0pt}{0pt}{*1}
\titleformat{\subsubsection}{\bfseries\sffamily\normalsize}{\thethesubsection.~}{0pt}{}
\titlespacing{\subsubsection}{0pt}{0pt}{*0}

   \@ifundefined{showcaptionsetup}{}{%
    \PassOptionsToPackage{caption=false}{subfig}}
   \usepackage{subfig}

 \makeatother

\title{Remarkable Hydrogen Storage on Beryllium Oxide Clusters: First Principles Calculations}

\author{Ravindra Shinde}
\affiliation{Department of Physics, Indian Institute of Technology Bombay, Mumbai, Maharashtra 400076, INDIA.}
\email{ravindra.shinde@iitb.ac.in}
  \phone{+91 (0)22 25764558}
  \fax{+91 (0)22 25767552}

\author{Meenakshi Tayade}
\affiliation{Department of Chemistry, Institute of Chemical Technology, Mumbai, Maharashtra 400019, INDIA}
\email{mkstayade@gmail.com}


\begin{document}
% \linenumbers
\noindent{\color{JPCCBlue}{\rule{\textwidth}{0.5pt}}}
\begin{abstract}
Since the current transportation sector is the largest consumer of oil, and subsequently responsible for major air pollutants, 
it is inevitable to use alternative renewable sources of energies for vehicular applications. The hydrogen energy seems 
to be a promising candidate. To explore the possibility of achieving a solid-state high-capacity storage of hydrogen for onboard applications, 
we have performed first principles density functional theoretical calculations of hydrogen storage properties of beryllium 
oxide clusters (BeO)$_{n}$ (n=2 -- 8). We observed that polar BeO bond is responsible for H$_{2}$ adsorption. The problem 
of cohesion of beryllium atoms does not arise, as they are an integral part of BeO clusters. The (BeO)$_{n}$ (n=2 -- 8) adsorbs 
8--12 H$_{2}$ molecules with an adsorption energy in the desirable range of reversible hydrogen storage. The gravimetric 
density of H$_{2}$ adsorbed on BeO clusters meets the ultimate 7.5 wt \% limit, recommended for onboard practical applications.
 In conclusion,  beryllium oxide clusters exhibit a remarkable solid-state hydrogen storage.
\end{abstract}
\noindent{\color{JPCCBlue}{\rule{\textwidth}{0.5pt}}}

\section{\label{sec:introduction}INTRODUCTION}
  Hydrogen energy has emerged as clean, highly efficient and eco-friendly option to fossil fuels \cite{jacs-mof,
jcat-spilover-mof,jacs-prussian-blue, admat-metal-inorganic-nanostructures, jacs-mof-spillover1, jacs-mof-spillover2, jpcb-catalyst-mgh2,
ijhe-h-gen-nabh2}. 
Hydrogen fuel delivers a very high energy per unit mass as compared to other fuels. However, the utilizable energy per unit volume is very low.
 Cryogenic liquid hydrogen or high-pressure hydrogen cylinder can achieve high volumetric densities. Such a storage may not be the safe 
option for vehicular applications. For practical onboard applications, 
hydrogen needs to be stored at optimum pressures and temperatures with easy reversibility and fast kinetics\cite{jacs-ti-on-c60,langmuir-adsorption}. 
The use of hydrogen as a fuel is hindered by the lack of economical methods of its storage. A solid-state storage of hydrogen using 
adsorption/desorption mechanism received a great deal of attention recently \cite{jacs-ti-on-c60, jacs-li-on-fullerene, li-on-bc3-jena-jpcc, 
jacs-mof,jcat-spilover-mof,jacs-prussian-blue, admat-metal-inorganic-nanostructures, jacs-mof-spillover1, 
jacs-mof-spillover2,jpcb-catalyst-mgh2,ijhe-h-gen-nabh2,langmuir-adsorption}.

Hydrogen, in molecular form tries to physisorb on the substrate surface rather weakly, and
in atomic form it chemisorbs strongly with substrate material. In the former case, a low hydrogen storage is achieved with hydrogen 
desorbing easily, and in the latter, a high storage is achieved, but desorption of H$_{2}$ is more difficult \cite{jacs-ti-on-c60,jacs-li-on-fullerene}. 
For an efficient and reversible hydrogen storage medium, it is advisable to have hydrogen adsorbed in molecular form with adsorption energy in the 
range of 0.1 to 0.4 eV/H$_{2}$ \cite{langmuir-adsorption, epl-li-doped-graphane,jacs-ti-on-c60,jacs-li-on-fullerene}. 
A potential storage material of H$_{2}$ would have large surface area and low molecular weight. The US Department of Energy (DOE) has 
set an ultimate goal that the gravimetric density of H$_{2}$ should exceed 7.5 wt $\%$  \cite{doe}. 

A natural choice of adsorbent material would be 
materials containing light elements. There has been an enormous exploration of possibilities of having carbon based, boron based and other 
doped materials that can store a significant amount of hydrogen. For example, metal based hydrides have been 
studied extensively \cite{jacs-ti-on-c60, li-on-bc3-jena-jpcc, jacs-prussian-blue,ijhe-h-gen-nabh2,jacs-mg-mgh-clusters,
jacs-hydrogen-on-mg-clusters,rscadv-all-metal-aromatic-clusters,jpcc-carbon-foams}.
However, in this case the gravimetric weight percentage is quite low also the hydrogen tries to chemisorb on the transition metals. This needs 
higher operating temperatures to desorb hydrogen. Carbon based structures in various forms have also been studied widely. Carbon foams,
hydrogen filled carbon nanotubes, graphene, graphane sheets, boron-nitride fullerenes, etc. are few representative 
examples \cite{cnt-cpl-2002,swcnt-cpl-2003,carbon-materials-storage-apl, epl-li-doped-graphane,jpcc-ca-decorated-graphyne,
jacs-ti-on-c60,ijhe-carbon-doped-bn-fullerene,jacs-li-on-fullerene,jpca-be-cages, chemphys-mg-decorated-boron, 
admat-metal-inorganic-nanostructures,jpcc-spillover-graphene-boron,jpcs-boron-nitride,li-on-bc3-jena-jpcc}. 
However, pure carbon based materials are found to adsorb hydrogen very weakly \cite{cnt-cpl-2002,swcnt-cpl-2003,carbon-materials-storage-apl}.
Low operating temperatures prohibits practical use of the metal-organic framework (MOF) hydrogen adsorbents \cite{jacs-mof,jacs-mof-spillover1,
jacs-mof-spillover2,csr-mof-cof}. Transition metal decorated fullerenes have potential of 
high-capacity hydrogen storage, but cohesion of decorated transition metals is still a problem. The metal adsorption energy needs to be 
larger than the cohesive energy of transition metal atoms.
To bypass the problem of cohesion of metal atoms, adsorbing hydrogen directly on a substrate is preferred \cite{chemphys-li2o}.
 Hence, in this paper, we systematically investigate high-capacity hydrogen storage behavior of beryllium oxide clusters using first principles density
functional calculations. These clusters are studied extensively and are found to be very stable \cite{jcp-electronic-structure-beo}. 
Since hydrogen adsorbs directly on these clusters, the cohesion problem, is eliminated. Also, the base clusters consist of only light-weight 
elements, making hydrogen gravimetric density quite larger than those of transition metal based substrates. A correlation of hydrogen 
adsorption energies per H$_{2}$, maximum possible hydrogen wt \%, incremental binding energies with respect to the increasing size 
of the beryllium oxide cluster has been presented. This work may help in experimental realization of high-capacity, sustainable
 hydrogen fuels, which is a need of today's and coming generations.

The remainder of the paper is organised as follows. Section \ref{sec:computational} describes details of the first principles calculations, 
followed by Section \ref{sec:results}, in which results are presented and discussed. In Section \ref{sec:conclusions}, we present conclusions
and discuss future implementation.

\section{\label{sec:computational}COMPUTATIONAL DETAILS}
All first-principles density functional theoretical calculations were performed using \textsc{Gaussian 09} code \cite{gaussian09}. Geometries of both
bare and hydrogen adsorbed BeO clusters were optimized without any symmetry constraints using the gradient embedded
genetic algorithm (GEGA).\cite{genetic1,genetic2,genetic3} In this method, initial geometries of species in the population are randomly generated
in one, two, and three dimensions. The population size is taken as 5N, where N is the number of atoms in the cluster. All
structures in this initial population are optimized to the nearest local minimum using \textsc{Gaussian 09} code using a smaller basis set.
All successfully converged local minimum structures undergo breeding and mutations. Structures with low overall energies
are preferred for breeding with probabilities depending on their energies. Couples of parents are randomly selected for breeding
based on these probabilities. A random plane cuts through both parents. The resultant half structures are then joined, forming a
child; i.e., the part above the cutting plane of parent 1 is combined with the part below the cutting plane of parent 2,
keeping the number of atoms the same. Such structures are again optimized to the nearest local minimum. This random
selection of parents is continued until the population gets doubled, i.e., N initial parents plus N newborn children. These
2N structures are sorted according to their energies, and a new set of N low-lying structures is formed. If the lowest energy
structure in each iteration of breeding remains the lowest for 20 iterations, then the algorithm is said to be converged to the
global minimum. Structures with higher energies are selected for mutations. The structures of one-third population are
randomly kicked out of equilibrium and then converged to the nearest minimum. This prevents trapping of species in a
particular local minimum. The mutants are added to the population along with their precursors, and the breeding
iterations continue. After convergence, the lowest energy structure and various low-lying structures are optimized with
a larger basis set at a higher level of theory.

We used the $\omega$B97xD energy functional along with an allelectron 6-311+G(d, p) basis set for the final optimization. The
effect of van der Waals interactions was included explicitly by using the empirical correction scheme of Grimme (DFT
+D2)\cite{grimme}. On the global minimum structures of BeO clusters, a number of H$_{2}$ molecules were successively added, and the
structures were reoptimized. This procedure was repeated until hydrogen remained adsorbed or the adsorption energy fell
below 0.10 eV/H$_{2}$. This ensured that H$_{2}$ remains physisorbed and feasible for reversible adsorption operational conditions.

\section{\label{sec:results}RESULTS AND DISCUSSION}
In this section, we present the results of first principles calculations, organized as follows. Following subsection presents geometrical structures
and stability of BeO clusters, followed by electronic structure of bare and hydrogen-decorated BeO clusters. In the final subsection, we discuss
hydrogen storage ability of BeO clusters.
\subsection{\label{subsec:binding}Structures and binding energies}
  In order to make sure that the adsorbent BeO clusters are stable, we calculated binding energies using the following formula

\begin{equation}
 E_{b} = -\frac{[E(BeO)_{n} - n\times E(BeO)]}{n}
\end{equation}
where $E(BeO)_{n}$ and $E(BeO)$ represent the total energies of (BeO)$_{n}$ cluster and a single BeO molecule, respectively. 
If the value of E$_{b}$ is positive, it means that the cluster formation is exothermic and therefore stable. The various isomers
of a given cluster, which are metastable, have not been considered here because the global minimum structure is the most favoured
if produced experimentally. 

\noindent{\color{JPCCBlue}{\rule{\columnwidth}{1pt}}}
\begin{table}
% \centering
 \caption{\label{tab:binding}Binding energy (E$_{b}$) of (BeO$_n$) (n=1--8) clusters, successive difference in 
 binding energy $\Delta E_b$, number of hydrogen molecules adsorbed (N$_{H_{2}}$), hydrogen molecule adsorption 
weight-percentage, and the adsorption energy per H$_{2}$ (E$_a$). } 
\begin{tabular}{MMMMMM}
\rowcolor{tablegray}
n  & E$_{b}$ (eV) &  $\Delta E_b$ (eV) & N$_{H_{2}}$ & wt \% & E$_a$ (eV) \\[1.5ex]
 2 & 3.45 &  	   &  12 & 32.60 & 0.12 \\
 3 & 4.86 & 1.41  & 12 & 24.38 & 0.10 \\
 4 & 5.33 & 0.47  & 8 & 13.88	& 0.10 \\
 5 & 5.51 & 0.18  & 5 & 7.46 &  0.10 \\
 6 & 5.58 & 0.07 & 6 & 7.46 &  0.12 \\
 7 & 5.64 & 0.06  & 7  & 7.46	& 0.16 \\
 8 & 5.89 & 0.25  & 8 & 7.46 & 0.13 \\
\end{tabular}
\noindent{\color{JPCCBlue}{\rule{\columnwidth}{1pt}}}
\end{table}

The binding energy of the clusters ($E_{b}$) and increment in the binding energy $\Delta E_b=E_{b}(n) - E_{b}(n-1)$, considered in this study,
 are listed in Table \ref{tab:binding}. Clearly, the binding energy of the clusters increases monotonically while $\Delta E_b$ 
keeps on decreasing as the cluster grows. However, for the (BeO)$_{8}$ cluster, a sudden increase in the $\Delta E_b$ is observed. This is 
mainly because of increase in the coordination number of Be atoms as they now bond with a greater number of oxygen atoms, thereby increasing the 
 stability. From these results, we can conclude that: (1) the large clusters are relatively more stable and therefore favorable. (2) The increased 
number of Be -- O bonds in larger clusters will further help in adsorbing more hydrogen.

\subsection{\label{subsec:ele-struct}Electronic structure}

\begin{figure}
\noindent{\color{JPCCBlue}{\rule{\columnwidth}{1pt}}}
\centering
\subfloat[]{\includegraphics[height=2.8cm]{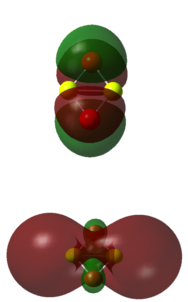}} \hfill
\subfloat[]{\includegraphics[height=2.8cm]{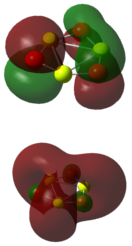}}\hfill
\subfloat[]{\includegraphics[height=2.8cm]{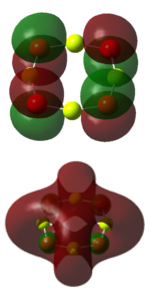}} \hfill
\subfloat[]{\includegraphics[height=2.8cm]{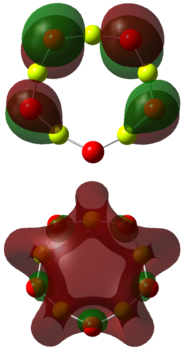}}  \\
\subfloat[]{\includegraphics[height=2.8cm]{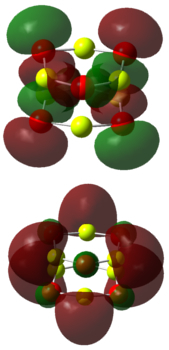}}\hfill
\subfloat[]{\includegraphics[height=2.8cm]{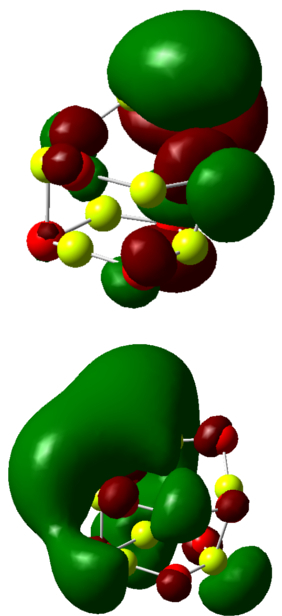}}\hfill
\subfloat[]{\includegraphics[height=2.8cm]{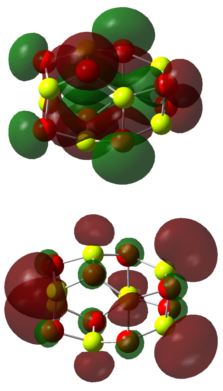}}
\caption{\label{fig:homo-lumo} HOMO and LUMO corresponding to global minimum structures of (BeO)$_{n}$ (n=2--8) clusters. In each subfigure,
top and bottom figures represent HOMO and LUMO respectively.}
\noindent{\color{JPCCBlue}{\rule{\columnwidth}{1pt}}}
\end{figure}

\begin{figure}
\noindent{\color{JPCCBlue}{\rule{\columnwidth}{1pt}}}
 \centering
 \includegraphics[width=3cm]{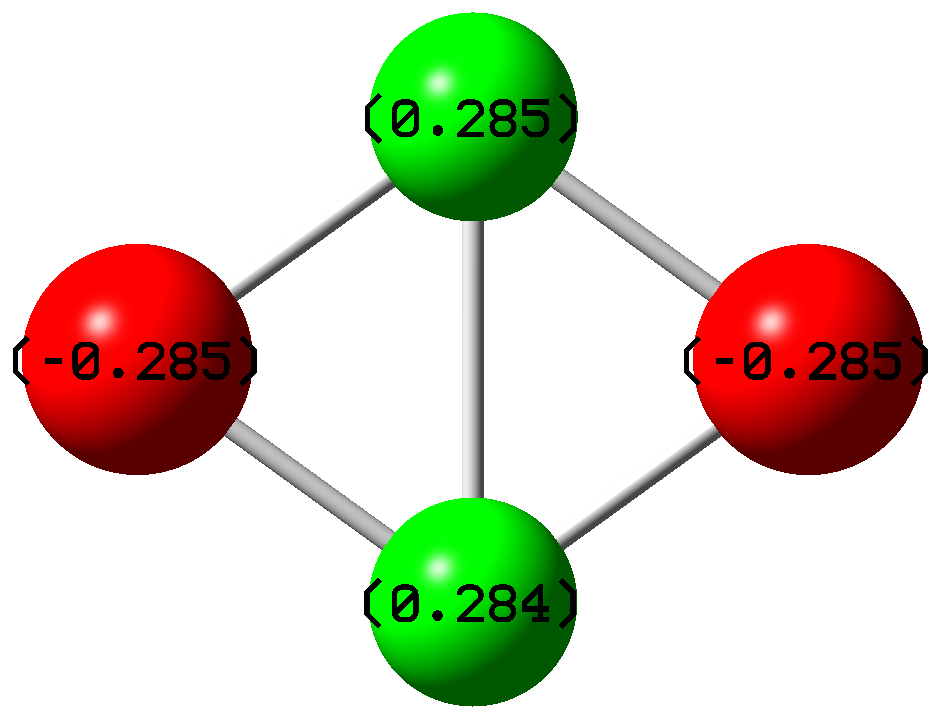}
 \caption{\label{fig:mulliken}Mulliken charges on individual atoms of (BeO)$_{2}$ cluster. Note that polarized Be -- O bond is responsible for hydrogen adsorption.}
\noindent{\color{JPCCBlue}{\rule{\columnwidth}{1pt}}}
\end{figure}

Highest occupied molecular orbitals (HOMO) and lowest unoccupied molecular orbitals (LUMO) of stable ground state geometries of 
(BeO)$_{n}$ (n=2--8) clusters are as shown in Fig. \ref{fig:homo-lumo}. HOMOs are mostly contributed from oxygen atoms, and the LUMOs
are by beryllium atoms. Depending upon the geometry, one beryllium atom can bind with two or more oxygen atoms. Clusters with beryllium 
bound to two oxygen atoms show enhanced hydrogen adsorption behavior as compared to those bound to three or more oxygen atoms.
 This is mainly because of the charge transfer mechanism that polarize the adsorbing hydrogen. For example, in the case of (BeO)$_{2}$ clusters,
 there is 0.14$e$ charge transfer from each beryllium atom to each oxygen atom (Fig. \ref{fig:mulliken}).
As a result, this Be--O bond becomes polar and produces a local electric field, enough to polarize hydrogen molecules and eventually binding them
to the cluster. 

\begin{figure}
\noindent{\color{JPCCBlue}{\rule{\columnwidth}{1pt}}}
 \centering
\centering
\subfloat[\label{subfig:esp-beo21}]{\includegraphics[width=3cm]{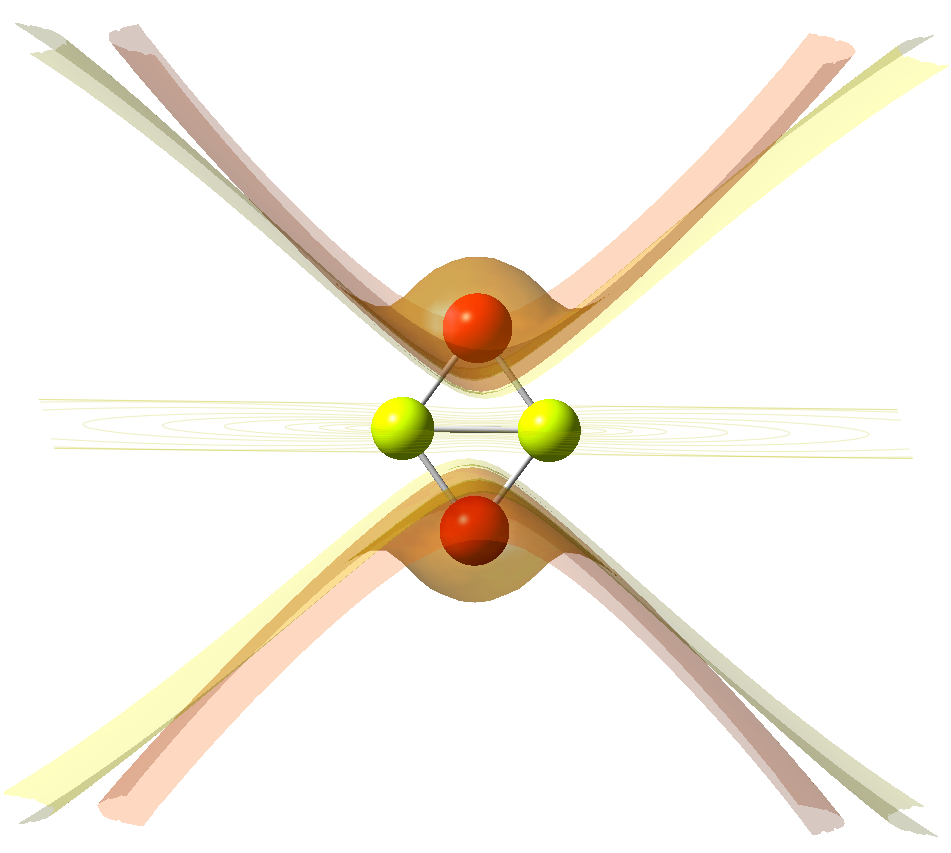}} \hspace{0.3cm}
\subfloat[\label{subfig:esp-beo21-with-h2}]{\includegraphics[width=4cm]{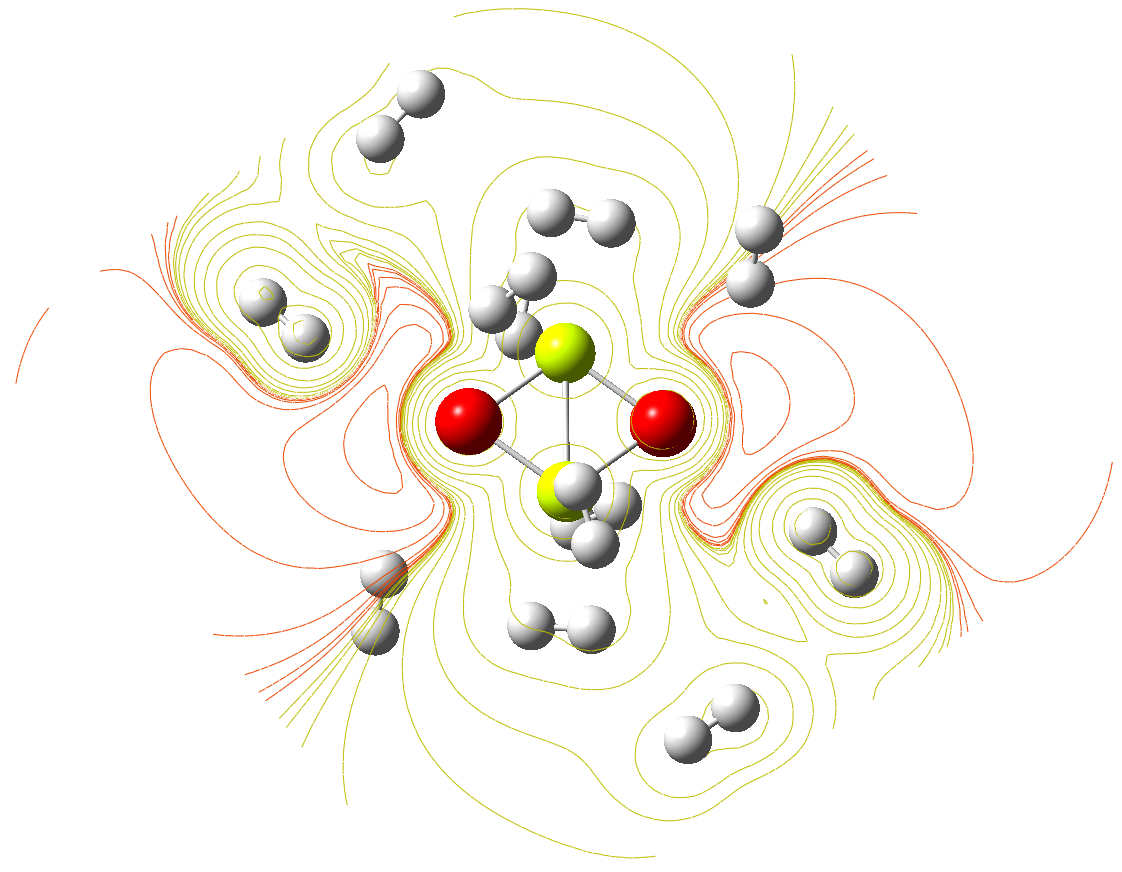}}
\caption{\label{fig:esp-curves}\protect\subref{subfig:esp-beo21} The electrostatic potential curves of (BeO)$_{2}$ cluster. \protect\subref{subfig:esp-beo21-with-h2}
The electrostatic potential contours of (BeO)$_{2}$ cluster with maximum hydrogen adsorbed in the plane containing (BeO)$_{2}$ cluster. 
Red color denotes the contribution from oxygen atoms and yellow color denotes beryllium.}
\noindent{\color{JPCCBlue}{\rule{\columnwidth}{1pt}}}
\end{figure}

A contour plot of electrostatic potential curves in the plane of BeO units can reveal favourable docking positions of hydrogen molecules.
Figure \ref{fig:esp-curves}\subref{subfig:esp-beo21} and \ref{fig:esp-curves}\subref{subfig:esp-beo21-with-h2} show that 
regions where oxygen is present; therefore, are avoided by hydrogen. This is consistent with Mulliken charge analysis that negatively charged oxygen
would repel hydrogen.

\subsection{\label{subsec:optimum_conditions}Optimum Conditions for Hydrogen storage}
The affinity of hydrogen toward an adsorbent should be strong enough to store a large amount of hydrogen at charging
pressures (about 30 bar) and weak enough to release most of the hydrogen at the discharging pressures (about 1.5 bar). In
the Langmuir isotherm approximation, the average energy of hydrogen adsorption between pressures P1 and P2 is given by\cite{langmuir-adsorption}:

\begin{equation}
 E_{a} = T\Delta S\gamma + \frac{RT}{2}ln \left( \frac{P_1 P_2}{P_0^2} \right)
\end{equation}

For variety of hydrogen adsorbents, the change in entropy is $\Delta S\gamma \approx$ 8R, with R being the ideal gas constant. At 298 K, with
P$_1$ = 30 bar, P$_2$ = 1.5 bar, and P$_0$ = 1 bar, the optimum hydrogen adsorption energy equals 0.15 eV\cite{langmuir-adsorption}. 
This means that, if E$_a$ is around 0.15 eV, then the hydrogen can be adsorbed and desorbed easily in this pressure range at room temperature.

\subsection{\label{subsec:storage}Hydrogen storage behavior of (BeO)$_{n}$ clusters}

\begin{figure*}
\centering 
\noindent{\color{JPCCBlue}{\rule{\columnwidth}{1pt}}} 
\subfloat[\label{subfig:beo21}]{\includegraphics[width=3.5cm]{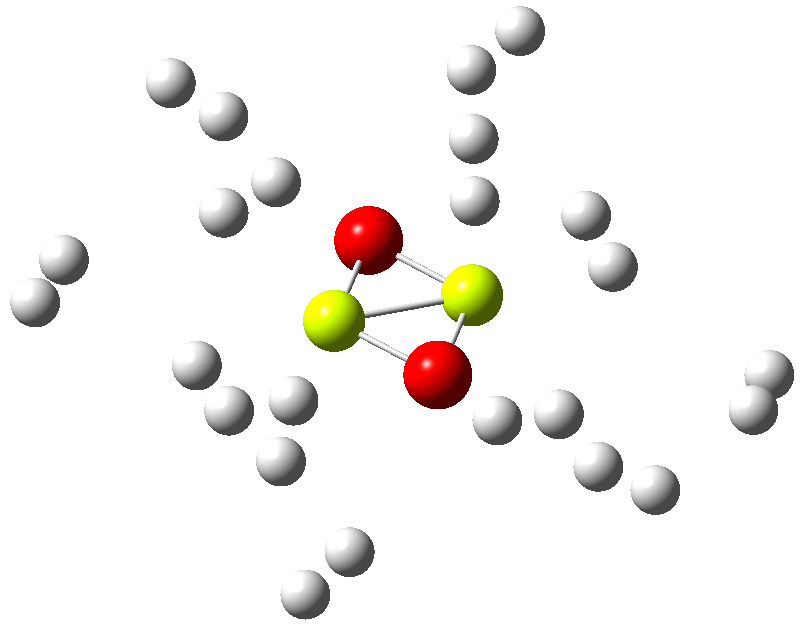}} \hfill
\subfloat[\label{subfig:beo31}]{\includegraphics[width=3.0cm]{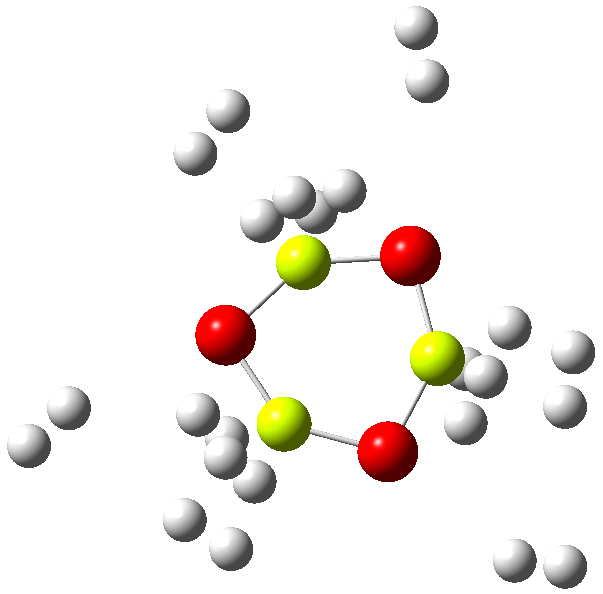}} \hfill
\subfloat[\label{subfig:beo41}]{\includegraphics[width=2.5cm]{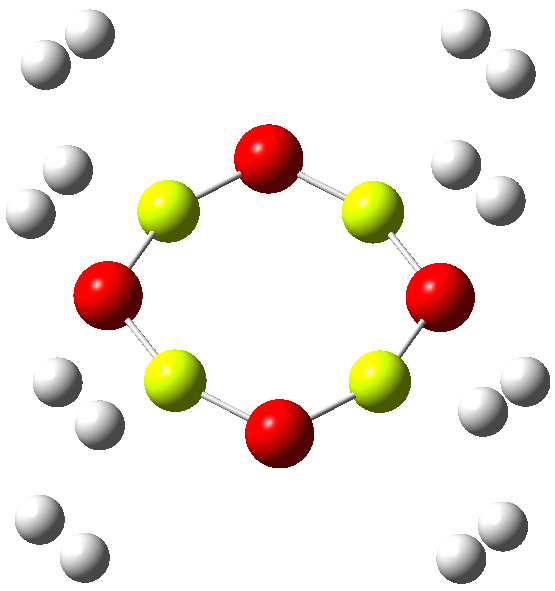}}  \hfill
\subfloat[\label{subfig:beo51}]{\includegraphics[width=3.0cm]{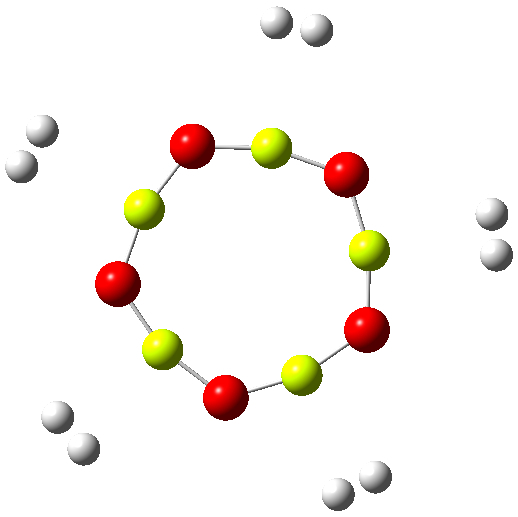}} \\
\subfloat[\label{subfig:beo61}]{\includegraphics[width=4.0cm]{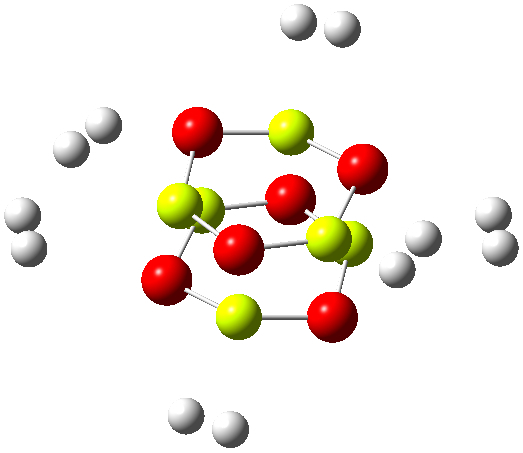}} \hfill
\subfloat[\label{subfig:beo71}]{\includegraphics[width=3.5cm]{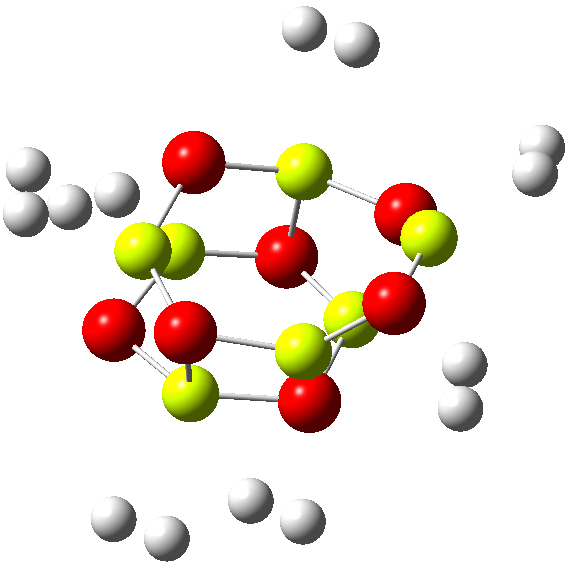}} \hfill
\subfloat[\label{subfig:beo81}]{\includegraphics[width=4.0cm]{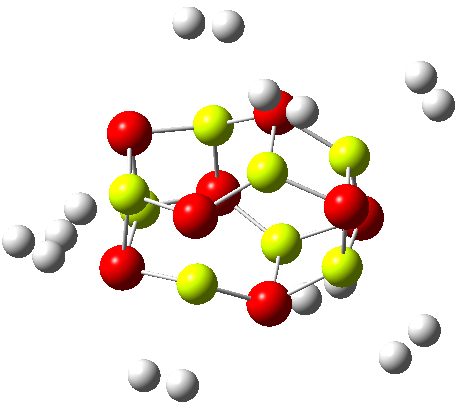}} \hfill
\caption{\label{fig:final_adsorbed}Geometry optimized global minimum structures of (BeO)$_{n}$ (n=2 -- 8) clusters 
with adsorbed hydrogen. Red, yellow and gray color denote oxygen, beryllium and hydrogen atoms respectively.}
\noindent{\color{JPCCBlue}{\rule{\columnwidth}{1pt}}}
\end{figure*}

To quantify the hydrogen uptake by (BeO)$_{n}$ clusters, we define hydrogen adsorption energy as,
\begin{equation}
  E_{a} = -\frac{[E_{(cluster+H_{2})} -E_{(cluster)}- m\times E_{(H_{2})}]}{m}
\end{equation}
where, $E_{(cluster+H_{2})}$ is the total energy of BeO cluster plus adsorbed hydrogens, $E_{(cluster)}$ is total energy of adsorbent
BeO cluster, $E_{(H_{2})}$ is the total energy of individual H$_{2}$ molecule and $E_{a}$ is hydrogen adsorption energy per H$_{2}$ molecule.
The number $m$ denotes the maximum number of H$_{2}$ molecule adsorbed on a given cluster.
For (BeO)$_{2}$, each Be atom adsorbs six H$_{2}$ as shown in Fig. \ref{fig:final_adsorbed}\subref{subfig:beo21}, which corresponds to
32.6 wt \% of hydrogen adsorption. The adsorption energy is around 0.12 eV/H$_{2}$, which is in the range, of 
reversible H$_{2}$ adsorption. The number of hydrogens adsorbed, hydrogen adsorption
energies and its gravimetric densities of all clusters studied here are listed in the Table \ref{tab:binding}.  The (BeO)$_{3}$ cluster can adsorb
12 H$_{2}$ molecules with 0.10 eV/H$_{2}$ adsorption energy obtaining 24 wt \% uptake (\emph{cf.} Fig. \ref{fig:final_adsorbed}\subref{subfig:beo31}). 
A slightly different orientation of adsorbed 
hydrogen can lead to 18 H$_{2}$ molecules with 0.093 eV/H$_{2}$ adsorption energy, further increasing hydrogen uptake to 32.6 wt \%. 
With 0.10 eV/H$_{2}$ adsorption energy, (BeO)$_{4}$ can adsorb upto 8 H$_{2}$ giving 14 wt \% gravimetric density (\emph{cf.} Fig. \ref{fig:final_adsorbed}\subref{subfig:beo41}). 
Clusters (BeO)$_{n}$, with n=5--8 (\emph{cf.} Fig. \ref{fig:final_adsorbed}\subref{subfig:beo51} -- \ref{fig:final_adsorbed}\subref{subfig:beo81}) 
adsorbs equal number of H$_{2}$ as that of Be giving rise to same gravimetric density of 7.46 wt \%. The 
hydrogen adsorption energy lies in the range of 0.10 -- 0.16 eV/H$_{2}$. It is noted that the number of hydrogens adsorbed decreases with an increase
in the cluster size. This is mainly due to the steric hindrance as well as reduced charge-transfer effect. Also, increased coordination number of Be
atoms in the cluster cause hydrogen adsorption energy to be slightly on the higher side. 

Since smaller clusters are favourable for high hydrogen uptake, these clusters can be used for devising efficient high-capacity 
hydrogen storage materials. Carbon foams, large nanotubes are found to adsorb H$_{2}$ with low wt \% 
\cite{jpcc-carbon-foams,carbon-cnt-low-adsorption,cnt-cpl-2002,swcnt-cpl-2003,carbon-materials-storage-apl}. 
The BeO clusters can be embedded in porous nanostructures, or can be supported on MOF, to achieve high volumetric hydrogen uptake.
\FloatBarrier
\section{\label{sec:conclusions}CONCLUSIONS AND OUTLOOKS}
Using first principles calculations we show that a remarkable high-capacity hydrogen storage can be achieved on (BeO)$_{n}$ clusters. The choice of using 
BeO clusters naturally avoids the aggregation problem usually found in metal-on-substrate type hydrogen adsorbents as the Be are the part 
of the substrate itself. The polar nature of Be--O bond - caused due to electronic charge transfer from Be to O - induces an electric 
field around positively charged Be atom which in turn polarizes H$_{2}$ and adsorbs them. The adsorption energies of H$_{2}$ 
are within 0.10-0.16 eV/H$_{2}$ which is a recommended range for reversible hydrogen physisorption under standard test conditions. This study 
may stimulate experimental efforts to check the claims of high-capacity, stable, reversible hydrogen adsorption, reported here.

\titleformat{\section}{\bfseries\sffamily\color{JPCCBlue}}{\thesection.~}{0pt}{\large$\blacksquare$\normalsize~}
\section*{AUTHOR INFORMATION}
\subsubsection*{Corresponding Author}
\noindent *E-mail: ravindra.shinde@iitb.ac.in
\subsubsection*{Notes} 
\noindent The author declares no competing financial interest.

\section*{ACKNOWLEDGMENTS} 
Author (R.S.) thanks the Council of Scientific and Industrial Research (CSIR), India, for research fellowship (09/087/(0600) 2010-EMR-I). 
Authors kindly acknowledge useful discussions with Dr. Garima Jindal.

\bibliography{hydrogen-storage-beo}

\begin{tocentry}
\includegraphics[width=8.5cm]{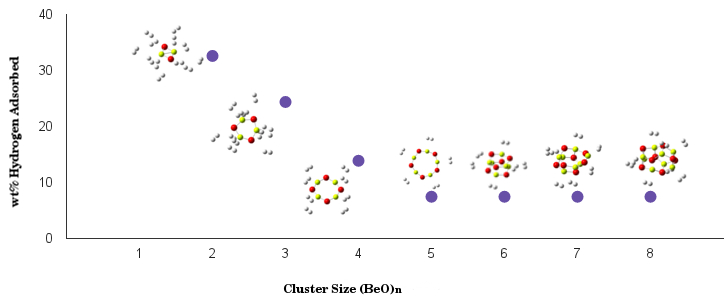}
\end{tocentry}
\end{document}